\documentclass{jfm}

\usepackage{graphicx}
\usepackage{natbib}
\usepackage{amsmath}
\usepackage{bm} 
\usepackage{pifont}
\usepackage{oubraces}
\usepackage{listings}
\usepackage{booktabs}
\usepackage{nicefrac}

\usepackage{tabularx}
\newcolumntype{Z}{>{\centering\arraybackslash}X}

\usepackage{mathrsfs}
\usepackage{microtype}
\usepackage{}
\usepackage{esdiff} 

\usepackage[pdfauthor={P.H. Trinh, S.K. Wilson, H.A. Stone}]{hyperref}

\ifCUPmtlplainloaded \else
  \checkfont{eurm10}
  \iffontfound
    \IfFileExists{upmath.sty}
      {\typeout{^^JFound AMS Euler Roman fonts on the system,
                   using the 'upmath' package.^^J}%
       \usepackage{upmath}}
      {\typeout{^^JFound AMS Euler Roman fonts on the system, but you
                   dont seem to have the}%
       \typeout{'upmath' package installed. JFM.cls can take advantage
                 of these fonts,^^Jif you use 'upmath' package.^^J}%
      }
  \else
  \fi
\fi

\ifCUPmtlplainloaded \else
  \checkfont{msam10}
  \iffontfound
    \IfFileExists{amssymb.sty}
      {\typeout{^^JFound AMS Symbol fonts on the system, using the
                'amssymb' package.^^J}%
       \usepackage{amssymb}%
         \let\leq=\leqslant
         \let\geq=\geqslant
      }{}
  \fi
\fi

\ifCUPmtlplainloaded \else
  \IfFileExists{amsbsy.sty}
    {\typeout{^^JFound the 'amsbsy' package on the system, using it.^^J}%
     \usepackage{amsbsy}}
    {}
\fi

\newsavebox{\astrutbox}
\sbox{\astrutbox}{\rule[-5pt]{0pt}{20pt}}

\title[An elastic plate on a thin viscous film]{An elastic plate on a thin viscous film}

\author[Trinh, Wilson, and Stone]
{P\ls H\ls I\ls L\ls I\ls P\ls P\ls E\ns H.\ns T\ls R\ls I\ls N\ls H$^{1}$,\ns
S\ls T\ls E\ls P\ls H\ls E\ls N\ns K.\ns W\ls I\ls L\ls S\ls O\ls N$^2$ \break
\and H\ls O\ls W\ls A\ls R\ls D\ns A.\ns S\ls T\ls O\ls N\ls E$^3$}

\affiliation{
$^1$ Oxford Centre for Industrial and Applied Mathematics (OCIAM), University of Oxford \\
Mathematical Institute, Andrew Wiles Building, Radcliffe Observatory Quarter, \\
Woodstock Road, Oxford, OX1 6GG \\[\affilskip]
$^2$ Department of Mathematics and Statistics, University of Strathclyde, \\
Livingstone Tower, 26 Richmond Street, Glasgow G1 1XH, UK \\[\affilskip]
$^3$ Department of Mechanical and Aerospace Engineering, Princeton University, \\
Princeton, New Jersey 08544, USA
}

\pubyear{XXXX}
\volume{YYY}
\pagerange{xxx--xxx}
\date{---}

\newcommand*{\de}{\operatorname{d\!}{}} 
\newcommand{\dd}[2]{\frac{\de#1}{\de#2}}
\newcommand{\pd}[2]{\frac{\partial#1}{\partial#2}}

\def\pa{p_{\hbox{a}}}
\def\hinf{h_{\infty}}

\newcommand{\rigidpaper}{TWS}
\newcommand*{\twotothree}{\text{II $\to$ III }}
\newcommand*{\twotoone}{\text{II $\to$ I }}

\newcommand*{\two}{\text{II}}
\newcommand*{\thr}{\text{III}}

\newcommand{\ti}[1]{{\tilde{#1}}}
\newcommand*{\Ca}{\mathrm{Ca}}
\newcommand*{\B}{\mathcal{B}}

 \newcommand*{\R}{\mathcal{R}}
 \newcommand*{\Oh}{\mathcal{O}}
 \newcommand*{\e}{\mathrm{e}}
 \newcommand*{\im}{\mathrm{i}}

 \newcommand*{\outie}{\textrm{outer}}
 \newcommand*{\lefty}{\textrm{left}}
 \newcommand*{\righty}{\textrm{right}}

\newcommand*{\cf}{\emph{c.f.}}
\newcommand*{\eg}{\emph{e.g.}}

\usepackage{tikz}
\newcommand*\mycirc[1]{%
  \tikz[baseline=-3pt] \draw[fill=white] (0,0) circle (4pt) node[scale=0.75] {\footnotesize\normalfont \textbf{#1}};
\!\!}

\usepackage{cleveref}

\begin{document}

\maketitle

\begin{abstract}
We consider the steady-state analysis of a pinned elastic plate lying on the free surface of a thin viscous fluid, forced by the motion of the bottom substrate moving at constant speed. A mathematical model incorporating elasticity, viscosity, surface tension, and pressure forces is derived, and consists of a third-order Landau-Levich equation for the thin film, and a fifth-order beam equation for the plate. A numerical and asymptotic analysis is presented in the relevant limits of the elasticity and Capillary numbers. We demonstrate the emergence of boundary-layer effects near the ends of the plate, which are likely to be a generic phenomenon for singularly perturbed elastocapillary problems.
\end{abstract}

\begin{keywords}
thin film, free-surface flows, fluid-structure interaction, elastocapillary flows
\end{keywords}

\section{Introduction}
\label{sec:intro}

\begin{figure} \centering
\includegraphics[width=1.0\textwidth]{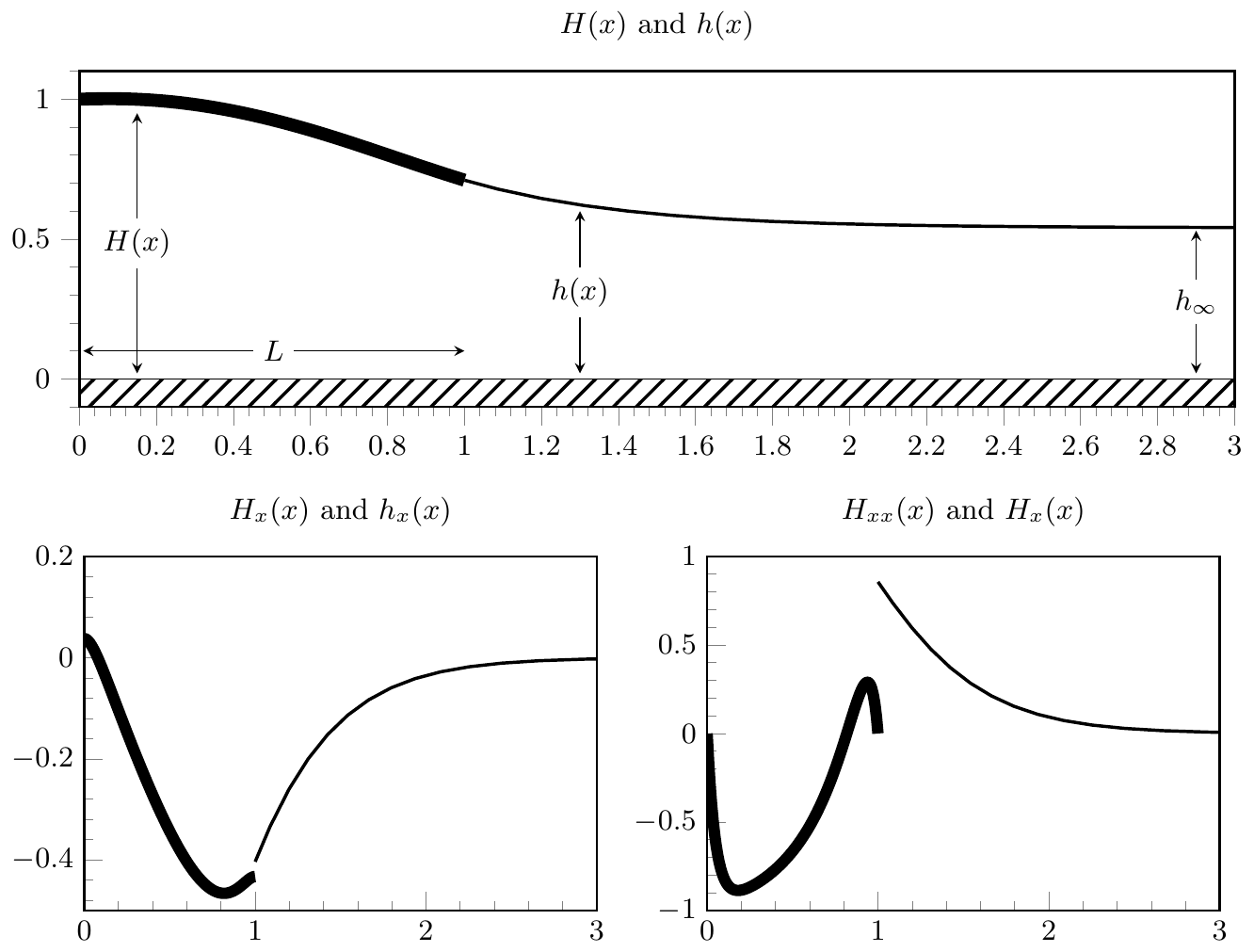}
\caption{The elastic plate (thick) and fluid free surface (thin) showing the non-trivial nature of the high derivatives. The function values, $H(x)$ and $h(x)$, are shown on top, and the first and second derivatives shown bottom left and bottom right. The numerical solutions correspond to $\delta = 1$, $\B = 0.3$, $p_0 = 1$. \label{fig:threeprofile}}
\end{figure}

We study the steady-state configuration of a finite elastic plate that lies on top of a thin viscous fluid. The plate is fixed and pinned to a fluid reservoir at one end, and the system is forced into motion by a bottom solid substrate moving to the right at constant speed (Figure \ref{fig:threeprofile}). The present problem is a paradigm for a wide range of elastocapillary and fluid-structure interaction flows, where competing effects of elasticity, viscosity, surface tension, and fluid pressure, can all play a role in determining the configuration of the plate. 

In a companion work by Trinh, Wilson, and Stone (\citeyear{trinh_2014a}) (henceforth written as \rigidpaper{}), an analogous study was presented for a similar situation involving a rigid plate (notably a pinned or free-floating plate on a thin film). There, it was noted by the authors that, even for this apparently simple scenario, a variety of interesting analytical and numerical challenges arise. For instance, the governing nonlinear boundary value problem for the steady-state configuration can exhibit nonuniqueness, leading to a complicated bifurcation structure of the solution space. The \rigidpaper{} study also led to the realization that numerical solutions of these coupled fluid-structure elasticapillary problems can be rather difficult to obtain, and often require analytical guidelines established using asymptotic analysis (\emph{e.g.} in the limits of large or small Capillary numbers).

Our main task in this work is to extend the \rigidpaper{} study to allow for an elastic and flexible plate, and to explore the theoretical and numerical properties of the proposed mathematical model in different physical limits. In doing so, not only do we recover many of the inherent challenges that appear in the former work, but we shall also encounter new difficulties that arise due to singular effects from the elasticity. 



Let us introduce the mathematical model. In Sec.~\ref{sec:mathform}, we shall derive two equations that govern the behaviour of the plate-fluid system illustrated in Figure~\ref{fig:threeprofile}. The first equation is a well known as the Landau-Levich equation: a third-order differential equation that describes the thin film profile, $h(x)$, and is given by
\begin{subequations} \label{intro:2eq}
\begin{equation} \label{intro:LL}
\delta^3 \diff[3]{h}{x} = \frac{3(h_\infty-h)}{h^3},
\end{equation}
where $h_\infty$ is the downstream height, and the non-dimensional quantity, $\delta^3$, is an inverse Capillary number [defined in \eqref{delta}] representing the balance between surface tension and viscosity. 

The second equation is a fifth-order Landau-Levich-\emph{like} equation that describes the profile, $H(x)$, of the elastic plate,
\begin{equation} \label{intro:elastic}
\B^5 \diff[5]{H}{x} - \delta^3 \diff[3]{H}{x} = \frac{6(H - 2h_\infty)}{H^3},
\end{equation}
\end{subequations}
and is effectively a linear beam equation with additional terms incorporating tension and pressure effects due to the underlying fluid. The quantity $\B^5$ is a non-dimensional parameter representing the balance between plate rigidity and viscous effects [defined in \eqref{defineB}]. The two equations, \eqref{intro:LL} and \eqref{intro:elastic}, are coupled together at the transition point, $x = L$, through continuity of the heights, as well as continuity of moment, shear, and pressure forces. 

The subtleties inherent in the study of the coupled boundary value problem can be seen in the three subplots of Figure~\ref{fig:threeprofile}, were we show a typical configuration at small values of the bending parameter (here $\B = 0.3$). In this example, we see that although the combined height profiles, $\{H(x), h(x)\}$, are well behaved, their derivatives are not. Notably, in the first derivative, there is a jump condition at the end of the plate which incorporates the shear stresses, while for the second derivative we see the formation of boundary layers near the ends, $x = 0$ and $x = L$. In Sec.~\ref{sec:Blimit} we shall present a matched asymptotic analysis of the $\B\to 0, \infty$ limits, and a similar asymptotic analysis is performed in Sec.~\ref{sec:deltlimit} for the $\delta \to 0, \infty$ limits.

\subsection{Background}

The Landau-Levich equation \eqref{intro:LL} is a canonical equation in the study of coating and draining problems, and arises from consideration of steady-state solutions of thin film dynamics. It was originally derived by \cite{landau_1942} and \cite{derjaguin_1943} to describe the thin film coating a plate withdrawn from a bath, but variants of the same equation occur in many other contexts, including the propagation of long gas bubbles in a channel \citep{bretherton_1961}, the coating the inside of a rotating cylinder \citep{ashmore_2003}, and the coating of fibres \citep{quere_1999}. We refer the readers to reviews by \cite{oron_1997}, \cite{stone_2010}, and \cite{de-gennes_2004}. 

In addition to their wide physical applicability, the Landau-Levich problem and its variations contain many challenging theoretical and numerical elements due to the nonlinearity of the differential equation. For instance, in studies of draining problems over a dry substrate, numerical works by \emph{e.g.} \cite{tuck_1990} and \cite{moriarty_1991} have emphasized the sensitivity of obtaining convergent steady-state solutions depending on the chosen numerical scheme. Still today, the full solution space of the Landau-Levich equation, subject to different boundary conditions, is not well understood due to the non-uniqueness of solutions \citep{snoeijer_2008, benilov_2010, ren_2014}.

Our motivation for studying the fifth-order Landau-Levich-\emph{like} equation \eqref{intro:elastic} stems from the recent interest in elastocapillary problems, where the goal is to better understand the interaction between an elastic solid and a fluid subject to the combined effects of viscosity and surface tension. Studies in this vein include the wetting of fibrous material (\eg{} \citealt{bico_2004}, \citealt{duprat_2012}, \citealt{taroni_2012}, and \citealt{singh_2014}), the buckling of floating elastic sheets (\eg{} \citealt{hosoi_2004}, \citealt{audoly_2011}, and \citealt{wagner_2011}), and the elastic drag-out problem (\eg{} \citealt{dixit_2013}). 

More directly related to the pinned plate-fluid system we study in this paper are the works on blade coating, which were summarized in Table 1 of the companion work \rigidpaper{}. Previous work by, for example, \cite{pranckh_1990}, \cite{iliopoulos_2005}, \cite{giacomin_2012}, \cite{seiwert_2013}, proposed mathematical models for such systems, but simplifying assumptions were often used (\eg{} neglecting surface tension, or simplified elasticity equations), which may not be valid in the regimes explored in our previous work. 

Lastly, we emphasize that while the pinned system proposed in this work is a very particular blade-coating setup, we expect that many of the mathematical properties of our study, notably the boundary-layer phenomena exhibited in Figure~\ref{fig:threeprofile}(c) will be widely applicable for fluid-elastic systems studied in their singular limits. These concerns were particularly highlighted in the previous work \rigidpaper{}, where we noted that in previous studies (\emph{e.g.} \citealt{moriarty_1996, conway_1983}) where free-floating problems were studied, authors were often unable to obtain solutions due to the complexity of the solution space, and singular perturbation effects. The solution of such fluid-structure interaction problems can be a delicate affair. 

\section{Mathematical formulation} \label{sec:mathform}

\noindent Let us consider steady two-dimensional flow of a thin film of Newtonian fluid with constant density $\rho$, viscosity $\mu$, and surface tension $\gamma$. The film lies on top of an rigid horizontal substrate, located at $\tilde{z} = 0$, that moves in the positive $\tilde{x}$ direction with constant speed $U$. The free boundary of the liquid will be composed of two parts: first an elastic plate of projected length $L$ lies along
\begin{equation}
\tilde{z} = \tilde{H}(\tilde{x}) \quad \text{for} \quad 0 < \tilde{x} < L,
\end{equation}
and second, we assume an uncovered free surface given by $\tilde{z} = \tilde{h}(\tilde{x})$ for $\tilde{x} > L$. In this work, we will consider the case that the left edge of the plate is fixed at a height $\tilde{H}(0) = \tilde{H}_0$, and is hinged and free to rotate (Figure~\ref{fig:threeprofile}). As we demonstrated in \cite{trinh_2014a} it is possible to study other variations of the boundary conditions (\eg{} a fixed plate angle at $\tilde{x} = 0$, a second free surface for $\tilde{x} < 0$, and so forth) and indeed these may introduced further subtleties. 

The pressure and velocity of the fluid are denoted by $\tilde{p} = \tilde{p}(\tilde{x})$ and $\mathbf{\tilde{u}} = (\tilde{u}(\tilde{x},\tilde{z}), \tilde{w}(\tilde{x},\tilde{z}))$, respectively, and we assume that the atmosphere above the plate and the free surface is inviscid and at a uniform atmospheric pressure denoted by $\pa$. We non-dimensionalise the variables according to
\begin{equation} \label{nondimen}
\begin{gathered}
\tilde{x} = L x, \quad
\tilde{z} = \epsilon L z, \quad
\ti{H} = \epsilon L H, \quad 
\tilde{h} = \epsilon L h, \\
\tilde{u} = U u, \quad
\tilde{w} = \epsilon U w, \quad
\tilde{p} - \pa = \frac{\mu U}{\epsilon^2 L} p,
\end{gathered}
\end{equation}
where $\epsilon = \ti{H}_0/L \ll 1$ is the aspect ratio of the film with respect to the height of the plate.

To derive the equations governing the free surfaces, $z = h(x)$, we apply the following boundary conditions on the solid and free surfaces:
\begin{subequations}
\begin{alignat}{4}
\text{no slip and no penetration on the substrate:} \quad &
(u,w) = (1,0) \quad & &
\text{on} \quad z = 0,
\label{noslipsubstrate} \\
\text{no slip and no penetration on the plate:} \quad &
(u,w) = (0,0) \quad & &
\text{on} \quad z = H(x),
\label{noslipplate} \\
\text{normal stress on the free surface:} \quad &
p = -\delta^3 h_{xx} \quad & &
\text{on} \quad z = h(x),
\label{p} \\
\text{tangential stress on the free surface:} \quad &
u_z = 0 \quad & &
\text{on} \quad z = h(x),
\label{tangential}
\end{alignat}
\end{subequations}
where we have defined an inverse capillary number
\begin{equation} \label{delta}
\textrm{Inverse Capillary number} = \delta^3 = \frac{\gamma \epsilon^3}{\mu U} = \frac{1}{\Ca}.
\end{equation}

Lubrication theory now allows us to derive the Reynolds equation for the pressure gradient, 
\begin{equation} \label{reynolds}
p_x = \frac{3(h - h_\infty)}{h^3},
\end{equation}
where $h \to \hinf$ as $x \to \pm \infty$. Substituting this result into the normal surface stress condition yields the Landau-Levich equation,
\begin{equation} \label{LandauLevich}
\delta^3 h_{xxx} = \frac{3(h_\infty-h)}{h^3},
\end{equation}
which governs the height of the free-surface. 

\subsection{The elastic plate and boundary conditions}

\noindent The equation for the fluid free surface, \eqref{LandauLevich} must be coupled to an equation governing the elastic plate, $z = H(x)$, which we now derive. First, let us consider the dimensional force per unit area, $\ti{\bm{F}} = [\ti{F}_1, \ti{F}_2]$, exerted by the fluid on the plate. Assuming a weakly tilted plate, whose outward normal is $\ti{\bm{n}} = [-\ti{H}_\ti{x}, -1]$, then we have the horizontal and vertical forces
\begin{equation}
\ti{F}_1 = -\ti{p} \dd{\ti{H}}{\ti{x}} + \mu \frac{\partial \ti{u}}{\partial \ti{x}} \qquad \text{and} \qquad
\ti{F}_2 = \ti{p},
\end{equation}
on the elastic plate. Let $\ti{T}$ be the tension of the plate, and by balancing the horizontal forces we have $\ti{T} = \ti{F}_1$. Consideration of the surface tension, $\gamma$, at the end of the plate suggests a re-scaling of $\ti{T} = \gamma T$, and we then have 
\begin{equation}
\frac{\partial \ti{T}}{\partial \ti{x}} 
\sim \frac{\mu U}{\epsilon \sigma} \left[ -\ti{p}\dd{\ti{H}}{\ti{x}} 
+ \pd{\ti{u}}{\ti{z}} \right] 
= \frac{\epsilon^2}{\delta^3} \left[ -\ti{p}\dd{\ti{H}}{\ti{x}} + \pd{\ti{u}}{\ti{z}} \right],   
\end{equation}
where we have used \eqref{nondimen} and \eqref{delta}. We stipulate that, for a given inverse Capillary number, $\delta^3$, the aspect ratio is such that $\epsilon^3 \ll \delta^3 = \Ca^{-1}$. Under this assumption, we would then have that the tension is constant, and thus the edge balance with surface tension implies $T \sim 1$ as a leading-order approximation. 

Next, we have the dimensional form of the normal stress balance given by [\cf{} eqn (4.4.1) in \cite{howell_2009}]
\begin{equation} \label{normbal}
\pd{\ti{N}}{\ti{x}} + \ti{T} \dd{^2 \ti{H}}{\ti{x}^2} + \ti{p}(\ti{x}) = 0,  
\end{equation}
where $\ti{N}$ is the transverse shear force. Applying conservation of angular momentum to a small surface element, we obtain the relation 
\begin{equation} \label{MNrelate}
  \dd{\ti{M}}{\ti{x}} = \ti{N}(\ti{x}),
\end{equation}
between the moment, $\ti{M} = \ti{M}(\ti{x})$, and the shear force. For small displacements, we assume the constitutive relation 
\begin{equation} \label{NHconst}
\ti{M} = -\mathrm{B} \dd{^2\ti{H}}{\ti{x}^2},
\end{equation}
where $\mathrm{B} = \mathrm{E}\mathrm{I}$ is the bending stiffness, as related to the elastic modulus, $\mathrm{E}$, and area moment of inertia, $\mathrm{I}$. We now differentiate \eqref{normbal}, substitute the constitutive relation \eqref{NHconst}, Reynolds equation \eqref{reynolds}, $\ti{T} = \gamma$, and non-dimensionalizations to obtain the final equation governing the elastic plate,
\begin{equation} \label{elasticeq}
\B^5 \dd{^5 H}{x^5} - \delta^3 \dd{^3 H}{x^3} = \R(H; h_\infty),
\end{equation}
where we have introduced the function
\begin{equation}
\R(H; h_\infty) = \frac{6(H - 2h_\infty)}{H^3},
\end{equation}
and the non-dimensional parameter
\begin{equation} \label{defineB}
\textrm{Elasticity number} = \B^5 = \frac{\mathrm{B} \epsilon^3}{\mu U L^2},
\end{equation}
measuring the balance between the effects of bending stiffness and viscosity.

Turning now to the boundary conditions at $x = 0$ and $x = 1$ for the plate, we recall that by the constitutive relation \eqref{NHconst}, the moment is proportional to the second derivative of $H$. Thus, the hinged and free ends satisfy 
\begin{subequations} \label{HHppbc}
\begin{gather}
H(0) = 1, \\
H_{xx}(0) = 0 = H_{xx}(1). \label{momentfree}
\end{gather}
\end{subequations}
In light of the fact that the free surface exerts a tangential force due to surface tension at the end of the plate, $x = 1$, the moment-free condition above may not be entirely obvious. We shall return to discuss this condition in Sec.~\ref{sec:govremark}.

\begin{figure}
\includegraphics{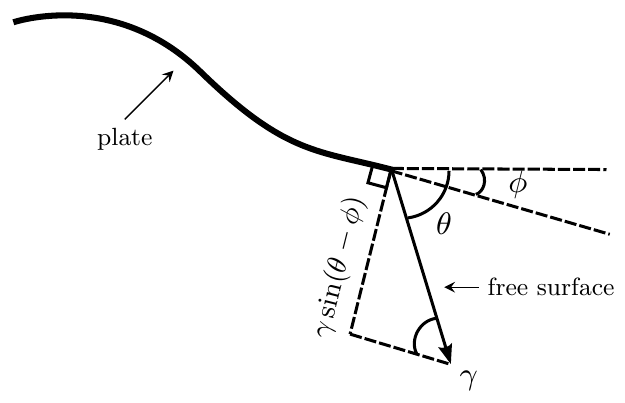} \hfill
\includegraphics{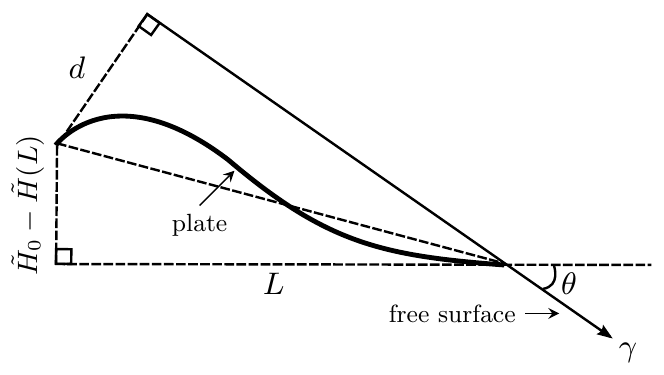} 
\caption{(Left) Quantities defined near the tip of the elastic plate and (right) illustration of the moment arm due to the surface tension forces \label{fig:tipzoom}}
\end{figure}

The next boundary condition follows from ensuring continuity of the normal shear forces at the end of the plate, $x = 1$. Examining Figure~\ref{fig:tipzoom}, we see that a balance of shear stress at the ends implies the dimensional balance,
\begin{equation}
  \ti{N}(L) + \ti{T} \Bigl[ \ti{H}_\ti{x}(L) - \ti{h}_\ti{x} \Bigr] = 0,
\end{equation}
and using \eqref{MNrelate} and \eqref{NHconst} gives the nondimensional form of the shear boundary condition:
\begin{equation} \label{shearbc}
  \B H_{xxx}(1) - \delta^3 \Bigl[ H'(1) - h'(1) \Bigr] = 0.
\end{equation}

We also require the pressure on the left to match the pressure of the reservoir, $p_0$, and the pressure on the right to match the Laplace-Young pressure, $-\delta^3 h_{xx}(1)$. Since the pressure underneath the elastic plate is given by \eqref{normbal} or alternatively the integral of \eqref{elasticeq}, we have 
\begin{subequations} \label{pressurebc}
\begin{align}
\B^5 H_{xxxx}(0) &= p_0, \label{pressurebc1} \\
\B^5 H_{xxxx}(1) &= -\delta^3 h_{xx}(1). \label{pressurebc2}
\end{align}
\end{subequations}

Finally, we require additional details in regards to the far field boundary conditions as $x \to \infty$. We linearize about the downstream height and write $h = h_\infty + \overline{h}$, where $\overline{h} \ll h_\infty$. Then according to a standard WKB analysis (see, for example, \citealt{tuck_1990}),
\begin{equation}
\overline{h} \sim
  C_1 \exp \left( -\frac{3^{1/3} x}{\delta\hinf} \right)
+ C_2 \exp \left(  \frac{3^{1/3}\e^{ \pi i/3} x}{\delta\hinf} \right)
+ C_3 \exp \left(  \frac{3^{1/3}\e^{-\pi i/3} x}{\delta\hinf} \right),
\label{eq:wkbmodes}
\end{equation}
as $x \to \infty$. The two exponentially growing modes, which represent capillary waves, are ruled out on physical grounds, so that $C_2 = 0$ and $C_3 = 0$, and leaving only $C_1$ to be determined. Thus, we see that the downstream condition $h \to h_\infty$ accounts for two boundary conditions in the third order Landau-Levich equation \eqref{LandauLevich}. This completes our derivation of the governing equations and boundary conditions. 


\begin{table}  \centering
\begin{tabular}{cp{5cm}p{6cm}}
\qquad Unknowns \qquad \qquad & Equations & Boundary conditions \\
$H(x)$, $h(x)$, $\hinf$ & 
$\B^5 H^{(5)} - \delta^3 H^{(3)} = 6(H-2h_\infty)/H^3$
\newline $\delta^3 h_{xxx}=3(\hinf-h)/h^3$
& 
[1] $H(0) = 1$
\newline [2] $H''(0) = 1$
\newline [3] $\B^5 H^{(4)}(0) = p_0$
\newline [4] $H(1) = h(1)$
\newline [5] $H''(1) = 0$
\newline [6] $\B^5 H'''(1) - \delta^3[H'(1) - h'(1)] = 0$
\newline [7] $\B^5  H^{(4)}(1) = \delta^3 h''(1)$
\newline [8,9] $h \to \hinf$ as $x \to \infty$\\
\end{tabular}
\caption{A summary of the equations and boundary conditions of the 9th order elastic pinned problem with eigenvalue $h_\infty$.}
\label{tab:pinned}
\end{table}

\subsection{Remarks on the governing equations} \label{sec:govremark}

A summary of the system of the two governing differential equations for the unknown plate profile, $H(x)$, fluid free surface, $h(x)$, and far-field height, $h = h_\infty$ is given by Table~\ref{tab:pinned}. In total, the system consists of the fifth-order plate equation and third-order fluid equation, so combined with the unknown eigenvalue, $h_\infty$, we expect nine boundary conditions. These nine boundary conditions consist of: [1, 2, 3] fixed height, zero moment, and reservoir pressure at $x = 0$; [4, 5, 6, 7] continuity of height, zero moment, continuity of shear and pressure at $x = 1$; [8, 9] far field conditions as $x \to \infty$. 

Although we chose to derive the governing equations and boundary conditions using local force and momentum balances, we note that it is also possible to use a variational approach. Using the small deflection and thin film approximations, we see that the equilibrium position of the plate should extremize 
\begin{multline} \label{energy}
\mathcal{E} = \int_0^L \biggl\{ \underbrace{\tfrac{1}{2} \mathrm{B} \tilde{H}_{\tilde{x}\tilde{x}}^2 }_{\text{bending energy}}- \underbrace{\tilde{p}\tilde{H}}_{\text{pressure energy}} 
+ \underbrace{\ti{T} \biggl[ \sqrt{1 + \tilde{H}_{\tilde{x}}^2} - 1\biggr]}_{\text{tension energy}} \, \biggr\} \de x \\
+ \int_L^\infty \biggl\{ \underbrace{ \gamma \biggl[ \sqrt{1 + \tilde{h}_{\tilde{x}}^2} - 1\biggr]}_{\text{surface tension energy}} - \underbrace{\tilde{p}\tilde{h}}_{\text{pressure energy}} \, \biggr\} \de x. 
\end{multline}
It can then be verified that the same equations and boundary conditions in Table~\ref{tab:pinned} can be derived if we seek require that the variation $\de{\mathcal{E}}/\de{\alpha}$ is zero.

Finally, in the statement of the moment-free condition, we indicated that it is not obvious that \eqref{momentfree} guarantees that the elastic plate is in rotational equilibrium. Although the conditions emerge naturally from application of the variable principal above, we can directly verify the moment calculation about $x = 0$. This is given by a pressure contribution and a contribution from the surface tension forces,
\begin{equation}
\ti{M}_0 = \int_0^L \ti{x} \ti{p}(\ti{x}) \, \de{\ti{x}} + \gamma d,
\end{equation}
where $d$ is the moment arm distance---that is, the perpendicular distance from $(0, \ti{H}_0)$ to the line parallel to the tension force in Figure~\ref{fig:tipzoom}. For small deflections of the elastic plate, this distance is 
\begin{equation}
   d \sim L \Bigl[ \ti{h}_\ti{x}(L) + \frac{\ti{H}_0 - \ti{H}(L)}{L}\Bigr].
\end{equation}
Finally, substitution of the pressure and boundary conditions indeed verifies that $\ti{M}_0 = 0$. 

\begin{figure} \centering
\includegraphics[width=1.0\textwidth]{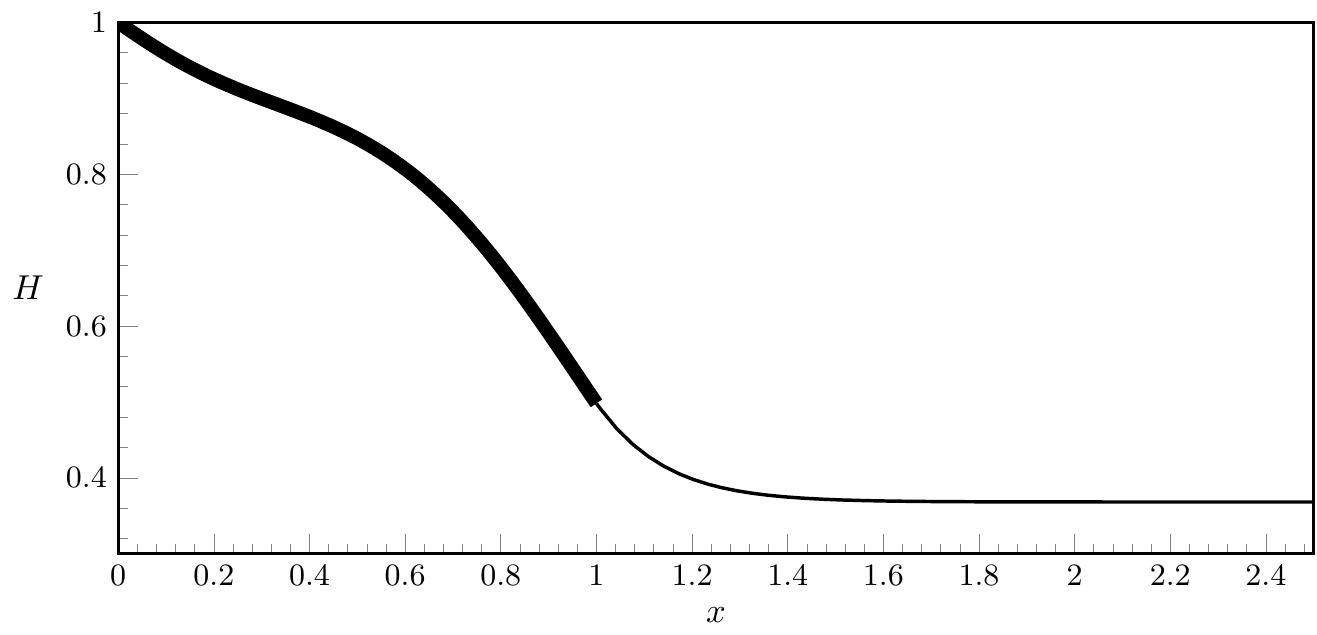}
\caption{A solution for the hinged configuration with $\B = 0.3$, $\delta = 0.5$, $p_0 = -0.5$ \label{fig:profile_p0neg}}
\end{figure}

\section{Numerical method}

We numerically solve the system in Table~\ref{tab:pinned} using finite differences and collocation, and explore the solution space using numerical continuation. It is often the case that finding an initial solution (or continuing a solution near singular limits) is difficult, and so we have either used the asymptotic predictions of \rigidpaper{} (for $\B = 0$) or the later asymptotic results of Secs.~\ref{sec:Blimit} and \ref{sec:deltlimit}, to obtain the preliminary solution.

For a given value of $\delta$, $\B$, and $p_0$, the fifth-order boundary value problem for $H(x)$ in \eqref{elasticeq} is solved with $h_\infty$ as an unknown eigenvalue. However, imposition of the six boundary conditions [1--3, 5--7] in Table~\ref{tab:pinned} requires values of $h'(1)$, and $h''(1)$. Thus within each iterate for solving the boundary-value problem, we must compute the Landau-Levich equation \eqref{LandauLevich}. To do this, we begin from an initial height of 
\begin{equation}
h = h_\infty + \varepsilon,
\end{equation}
where $\varepsilon$ is a small number ($\varepsilon = 10^{-12}$ in most computations). Using the exponentially decaying behaviour in \eqref{eq:wkbmodes} to provide values of the first and second derivatives, we solve the Landau-Levich equation in the negative $x$ direction as an initial value problem, stopping once $h = \hat{H}(1) \approx H(1)$ is reached (note that at this point, $H(1)$ is only known approximately as an intermediary value of the boundary-value iteration). Once the Landau-Levich equation has been solved, the approximate values of $h'(1)$ and $h''(1)$ can be collected, and the system is closed (with six boundary conditions for the fifth-order plate problem). An example of a numerical solution for positive pressure, $p_0 = 1$, is shown in Figure~\ref{fig:threeprofile}, and an example for negative pressure, $p_0 = -0.5$, is shown in Figure~\ref{fig:profile_p0neg}.

\section{Asymptotic analysis of large and small plate rigidity} \label{sec:Blimit}

In this section, we are interested in studying the system in Table~\ref{tab:pinned} for fixed values of the inverse Capillary number, $\delta$, and in the limit of large ($\B \to \infty$) and small ($\B \to 0$) plate rigidity. Although the limit of small rigidity has no analogue with our previous \rigidpaper{} work on rigid plates, we would expect that in the limit $\B \to \infty$, we are able to recover the result when the plate elasticity is ignored altogether. While this turns out to be the case, the more interesting phenomena is how the addition of non-zero (but small) elasticity affects the underlying equilibrium configuration. 

\begin{figure}\centering
\includegraphics{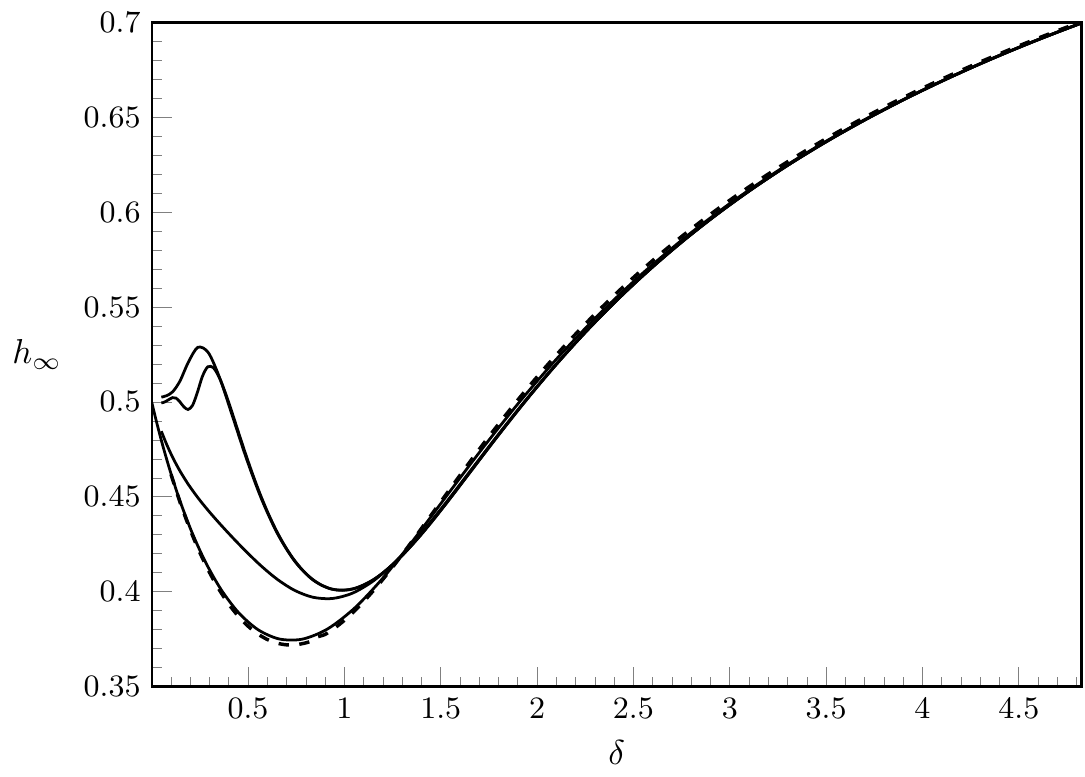}
\caption{Values of $h_\infty$ as a function of time for various values of $\B$. The reservoir pressure is $p_0 = 0$. From top to bottom, the solid curves are at $\B = 0.2, 0.1, 0.5, 1.0$. The dashed curve corresponds to $\B \to \infty$. \label{hinf_tautoinf}}
\end{figure}

\subsection{The rigid plate limit, $\B \to \infty$}

In the limit $\B \to\infty$, we assume there are no boundary layers (notably at $x = 0$ and $x = 1$), and expand the free surfaces as
\begin{equation} \label{Hseries_largeB}
h(x) = \sum_{n=0}^\infty \frac{h_n}{\B^{5n}} \qquad \text{and} \qquad H(x) = \sum_{n=0}^\infty \frac{H_n}{\B^{5n}}.
\end{equation}
At leading order in \eqref{elasticeq}, using $H_0(0) = 1$ and $H_0''(0) = 0$, we obtain the rigid plate,
\begin{equation} \label{H0}
H_0(x) = 1 + \alpha x,  
\end{equation}
where $\alpha$ is constant. What is somewhat curious is that in order to obtain the values of $\alpha$ and $h_\infty$, the moment and force balances must be applied at $\Oh(1/\B^5)$ and using the $H_1(x)$ correction. This is interesting because in the rigid formulation, the curvatures (and all higher derivatives) are identically zero, but for the elastic formulation, these higher-order derivatives become and essential for properly satisfying the boundary conditions. 

At $\Oh(1/\B^5)$ in \eqref{elasticeq}, the first-order solution satisfies 
\begin{equation} \label{H1}
H_1^{(5)} - \delta^3 H_0^{(3)} = \frac{6}{H_0^3} (H_0 - 2h_{\infty 0}),
\end{equation}
where we have also expanded the far-field height
\begin{equation}
 h_\infty = \sum_{n=0}^\infty \frac{h_{\infty n}}{\B^{5n}}.
\end{equation}
We integrate \eqref{H1} from $x = 0$ to $x = 1$, and use the moment free \eqref{momentfree} and pressure \eqref{pressurebc} boundary conditions to obtain a leading-order shear constraint, defined by
\begin{equation} \label{H1shear}
F_z^{(0)} = P_0 + \delta^3 h_{xx}(1) + \Bigl[6 I_2(1) - 12h_{\infty 0} I_3(1)\Bigr] = 0,
\end{equation}
after rearranging, where we have introduced the notation
\begin{equation}
I_k(x) = \int_0^x \frac{1}{H_0^k(x')} \, \de{x'}.  
\end{equation}

In fact, the shear condition \eqref{H1shear} was obtained in \cite{trinh_2014a} (their eqn (2.15)) but with our $H_0(x)$ replaced by the full $H(x)$ for the analogous study that ignores elasticity. Thus \eqref{H1shear} is a leading-order shear constraint, $F_z \sim F_z^{(0)}$. Since there are two remaining unknowns, $\alpha$ and $h_{\infty 0}$, to be determined at $\Oh(1)$, we expect to complement \eqref{H1shear} with an additional equation expressing a moment balance. We integrate \eqref{H1} three times and use the zero moment conditions \eqref{momentfree} to obtain after rearranging
\begin{multline} \label{prevert}
H_1^{(2)}(x)  - H_1^{(4)}(0)\biggl[\frac{x^2}{2} - x\biggr]  - x H_1^{(3)}(1) + \delta^3 \biggl[ H_0'(1) - \Bigl\{H_0(x) - H_0(0)\Bigr\}\biggr]\\
= \int_0^x \int_1^{x_2} \Bigl[6I_2(x_1) -12 h_{\infty 0}I_3(x_1)\Bigr] \, \de{x_1}\de{x_2}. 
\end{multline}
Setting $x = 1$ in \eqref{prevert}, reversing the order of integration, and applying the boundary conditions \eqref{shearbc}, \eqref{pressurebc}, and the rigid form \eqref{H0}, we obtain a leading-order moment constraint defined by
\begin{equation} \label{M00}
M_0^{(0)} = \frac{p_0}{2} + \delta^3 \Bigl[h_0'(1) - \alpha\Bigr] + \int_0^1 \Bigl[6x I_2(x_1) - 12 h_{\infty 0} x_1 I_3(x_1)\Bigr] \, \de{x_1} = 0,
\end{equation}
which matches with the moment about $x = 0$ from eqn (3.5) of \cite{trinh_2014a}. Our \eqref{M00} above, however, represents a leading-order constraint, $M_0 \sim M_0^{(0)}$ valid in the limit $\B \to \infty$. 

In summary, this demonstrates that the $\B \to \infty$ limit is nearly the same as the formulation where elasticity is entirely ignored, but the two formulations is that when elasticity is included and $1/\B \neq 0$, the plate is not quite flat, but is gently curved with $\Oh(1/\B^{5})$ curvature. The determination of the final two components, $\{\alpha, h_{\infty 0}\}$, requires the solution of the third-order Landau-Levich equation \eqref{LandauLevich}, subject to two infinity conditions ([8, 9] in Table~\ref{tab:pinned}), the continuity condition $h_0(1) = H_0(1) = 1 + \alpha$, and the two force conditions \eqref{H1shear} and \eqref{M00}. Although this can be done numerically, asymptotic formulae in the limits of $\delta \to 0$ and $\delta \to \infty$ were derived in \cite{trinh_2014a}. For instance, in the limit $\delta \to \infty$, 
\begin{equation}
\alpha \sim \frac{p_0 - 6}{3^{1/3} \delta^2} \quad \text{and} \quad
h_{\infty 0} \sim 1 + \frac{p_0 - 6}{3^{2/3}\delta},
\end{equation}
while the limit of $\delta \to 0$ is more complicated due to the existence of possibly more than one unique solution due to the effects of the reservoir pressure, $p_0$. 

In Figure~\ref{hinf_tautoinf}, we present the far-field thin film height, $h_\infty$, as a function of $\delta$ for values of $\B$ ranging from $\B = 0.1$ to $\B = 1$. It can be seen that by $\B = 1$, the curve is nearly identical to within visual accuracy to the $\B \to \infty$ limit computed using the full numerical solution of the rigid formulation. 

\subsection{The flexible plate limit, $\B \to 0$} \label{sec:B0}

We now study the limit in which the ratio between bending stiffness and viscous effects tends to zero. In fact, an asymptotic analysis of this regime turns out to be quite useful because the numerical solution of the governing boundary value problem is increasingly stiff as $\B \to 0$. We would like to expand the (outer) solution as 
\begin{equation}
  H(x) \sim H_\text{outer}(x) = \sum_{n=0}^\infty \B^n H_n,
\end{equation}
with similar expansions for $h_\infty \sim h_{\infty 0}$ and $h(x) \sim h_0(x)$. The leading-outer equation for the plate, from \eqref{elasticeq}, is
\begin{equation}
-\delta^3 H_0''' \sim \R(H_0; h_\infty),
\end{equation}
and so we need to provide four boundary conditions at $x = 0$ and $x = 1$. 

Somewhat surprising, though, is a rather complicated boundary layer structure in the singular limit. We will begin by considering the simpler case of zero imposed reservoir pressure, $p_0 = 0$, and then return to the case of general values of $p_0$ in Sec.~\ref{sec:p0notzero}. When $p_0 = 0$, it can be verified from the numerical profiles that there is no boundary layer at $x = 0$, and thus we can apply $H(0) = 1$ and $H''(0) = 0$. The remaining three conditions must be selected from amongst [3, 5, 6, 7] in Table~\ref{tab:pinned}, and our challenge is to determine which ones. As it turns out, the outer solution, valid away from $x = 1$, is coupled to inner solutions near $x = 1$ via three boundary layers. 

The full plate equation involves contributions from the plate rigidity, surface tension, and pressure. We write
\begin{equation} \label{elasticbalance}
\underbrace{\vphantom{[}\B^5 H^{(5)}}_{\mycirc{1}}
- \underbrace{\vphantom{[}\delta^3 H^{(3)}}_{\mycirc{2}}
= \underbrace{\vphantom{[}\R(H; h_\infty)}_{\mycirc{3}}.
\end{equation}
We define the following regions and asymptotic balances:

\mbox{}\par
\begin{center}
\begin{tabular}{r@{\hskip 0.8cm}>{$}l<{$}@{\hskip 0.8cm}rccc}
Outer & x - 1 = \Oh(1), & where \mycirc{2} & $\sim$ & \mycirc{3} \\[0.1cm]
Region I & x - 1 = \Oh(\B^{\alpha}), \ \alpha < 5/2 & where \mycirc{2} & $\gg$ & \mycirc{1} \\[0.1cm]
Region II &  x - 1 = \Oh(\B^{5/2}), & where \mycirc{1} & $\sim$ & \mycirc{2} \\[0.1cm]
Region III &  x - 1 = \Oh(\B^{\alpha}), \ \alpha > 5/2 & where \mycirc{1} & $\gg$ & \mycirc{2}
\end{tabular}
\end{center}
\mbox{}\par

%
In the outer region, surface tension balances the contribution from pressure forces. Moving towards the inner layer, and $x \to 1$, as the first boundary layer (Region I) is encountered, surface tension begins to dominate; in the middle boundary layer (Region II) surface tension and elasticity balance; and in the innermost layer (Region III), elasticity dominates. 

\begin{figure}\centering
\includegraphics[width=1.0\textwidth]{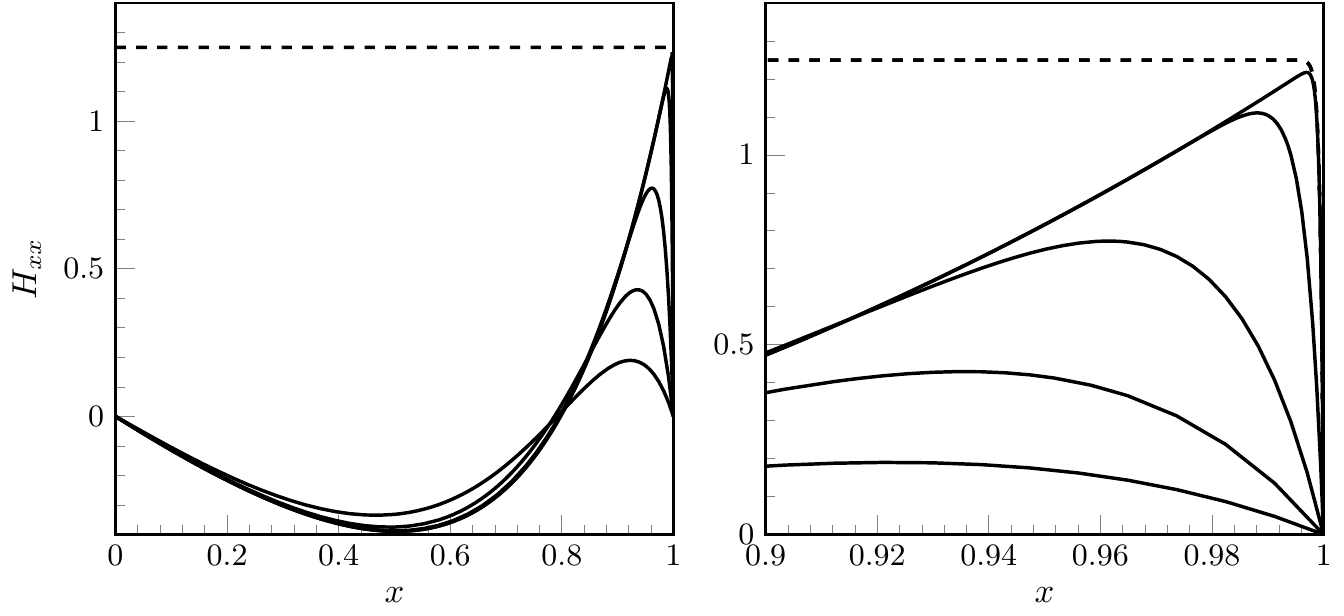}
\caption{Profiles of $H_{xx}$ for various values of $\tau$. All solutions as for $\delta = 1$ and $p_0 = 1$. Shown on the right is an enlargement of the region near $x = 1$. The dashed line is the asymptotic solution of Region II in \eqref{Hxx2}. From top to bottom on the right near $x = 1$, the solid curves correspond to $\tau = 0.05, 0.1, 0.2, 0.3, 0.4$. \label{fig:Hxxtau0}}
\end{figure}

Before we present the analysis, we present in Figure~\ref{fig:Hxxtau0} typical profiles of $H''(x)$ computed from the numerical solutions of the full system. The reason for examining the curvature of the plate (rather than $H$ or $H'$) is that the boundary layers only appear in the leading-order terms of the second and higher derivatives. Crucially our numerical intuition indicates that
\begin{equation} \label{keyassume}
  \text{$H(x)$, $H'(x)$, $H''(x)$ are bounded and non-zero as $\B \to 0$},
\end{equation}
for fixed values of $\delta$ and fixed $x \in (0,1)$.

Based on the figure, we see that although the zero-moment condition \eqref{momentfree} requires $H''(1) = 0$, the leading-order solution in the bulk has a curvature that approaches a non-zero value as $x \to 1$. Our principal task is to derive the effective value of the curvature to be applied to the outer solution, as marked by the dashed line in Figure~\ref{fig:Hxxtau0}.

\subsubsection{Region II: balancing elasticity with surface tension} \label{sec:region2}

We begin in Region II, where the elasticity effect \mycirc{1} balances the surface tension contribution \mycirc{2}. Solving the leading-order equation $\B^5 H^{(5)} \sim \delta^3 H^{(3)}$ gives
\begin{equation}
H^{\two}_{xxx} \sim A \exp \left[-\frac{\delta^{3/2}}{\B^{5/2}} (1-x)\right]
+ \hat{A} \exp \left[\frac{\delta^{3/2}}{\B^{5/2}} (1-x)\right],
\end{equation}
where $A$ and $\hat{A}$ are constants (possibly dependent on $\B$ and $\delta$). We intuit that $\hat{A} = 0$, as this suppresses the unmatcheable exponential growth. Integrating once yields
\begin{equation} \label{Hiixx1}
H^{\two}_{xx} \sim \frac{A\B^{5/2}}{\delta^{3/2}} \exp \left[-\frac{\delta^{3/2}}{\B^{5/2}} (1-x)\right]
+ C,
\end{equation} 
where $C$ is constant. However, note from the assumption \eqref{keyassume} that the curvature is bounded and non-zero, and so $A$ and $C = \Oh(\B^{-5/2})$. We therefore re-scale the constants and write \eqref{Hiixx1} as
\begin{equation} \label{Hiixx2}
H^{\two}_{xx} \sim \frac{\tilde{A}}{\delta^{3/2}} \exp \left[-\frac{\delta^{3/2}}{\B^{5/2}} (1-x)\right]
+ \tilde{C}.
\end{equation} 

In Region III, we assume the scaling $(1 - x) = \B^\alpha s$ for $\alpha > 5/2$ and $s = \Oh(1)$. Thus if we take $x$ from Region II to III, then within the exponential argument, $(1-x)/\B^{5/2} = \B^\alpha s/\B^{5/2}$, and this quantity must be small. Consequently, we may expand \eqref{Hiixx1} as follows:
\begin{equation}
H^{\twotothree}_{xx} \sim \left[\frac{\ti{A}}{\delta^{3/2}} - \frac{\ti{A}(1-x)}{\B^{5/2}} + \Oh\left(\frac{(1-x)^2}{\B^{10/2}}\right)\right] + \ti{C}. \label{2to3aa}  
\end{equation}

By the moment free boundary condition \eqref{momentfree}, we require $H''(1) = 0$. We will make the assumption (which can be verified \emph{a posteriori}) that this applies even for the inner limit of the Region II solution, and thus
\begin{equation} \label{Cval}
\ti{C} = -\frac{\ti{A}}{\delta^{3/2}}.
\end{equation}

Proceeding in the other direction, and expanding \eqref{Hiixx1} from Region II to Region I, we find that 
\begin{equation} \label{2to1aa}
H^\twotoone_{xx} \sim -\frac{\ti{A}}{\delta^{3/2}} + \Bigl[\text{exp. small in $(1-x) \delta^{3/2}/\B^{5/2}$}\Bigr],
\end{equation}
once we have made the substitution \eqref{Cval}. Thus the curvature is effectively constant to all algebraic orders, and it is precisely this constant, $\ti{C} = -\ti{A}/\delta^{3/2}$, which is shown as the dashed line in Figure~\ref{fig:Hxxtau0}. 

\subsubsection{Region I: surface tension dominates elasticity and pressure}

In Region I, we have $\delta^3 H_{xxx} \ll 1$, which gives $H^\text{I}_{xx} \sim \text{constant}$. This solution must match with \eqref{2to1aa}, so the constant must be equal to \eqref{Cval}, and thus
\begin{equation}
  H^{\two}_{xx} = -\frac{\ti{A}}{\delta^{3/2}}.
\end{equation}

\subsubsection{Region III: elasticity dominates surface tension and pressure}

Let us consider the solution in the innermost Region III, where elasticity effects from \mycirc{1} dominate both surface tension and pressure effects. In this case, we have $H^{(5)} \ll 1$, and so integrating
\begin{equation} \label{sol3a}
H^{\thr}_{xx} \sim \frac{\bar{A}}{2} (1-x)^2 + \bar{B}(1-x) + \bar{C}.
\end{equation}
By the moment free condition \eqref{momentfree}, $H_{xx}(1) = 0$, and thus we need $\bar{C} = 0$. Based on the $H^\twotothree$ limit of \eqref{2to3aa}, we argue that the solution in Region III cannot diverge quadratically as $x$ moves away from $x = 1$, and consequently, $\bar{A} = 0$. Then we match \eqref{2to3aa} and \eqref{sol3a} to obtain 
\begin{equation} \label{barB}
\bar{B} = -\frac{\tilde{A}}{\B^{5/2}}.  
\end{equation}

\subsubsection{Summary and final matching}

At this point, we have the following solutions:
\begin{subequations}
\begin{align}
H_{xx}^\text{I} &\sim -\frac{\tilde{A}}{\delta^{3/2}}, \label{H1aa} \\
H_{xx}^\text{II} &\sim \frac{\tilde{A}}{\delta^{3/2}} \exp \left[-\frac{\delta^{3/2}(1-x)}{\B^{5/2}} \right] -\frac{\tilde{A}}{\delta^{3/2}}, \label{Hxx2} \\
H_{xx}^\text{III} &\sim -\frac{\tilde{A}(1-x)}{\B^{5/2}}.
\end{align}
\end{subequations}

We now apply the remaining boundary conditions. First, since $\B^5 H'''(1) = \Oh(\B^{5/2})$ according to the solution in Region III (or II), the tension condition in \eqref{shearbc} reduces to 
\begin{equation}
H_x(1) \sim h_x(1),  
\end{equation}
at leading order. Thus, as $\B \to 0$, and the rigidity tends to zero, the plate and water surface should be tangential at contact. The pressure condition \eqref{pressurebc2}, applied to the inner limit of the solution in Region II \eqref{Hxx2} indicates that 
\begin{equation}
\B^5 H_{xxxx}(1) \sim \B^5 \frac{\ti{A}}{\delta^{3/2}} \left(\frac{\delta^{3/2}}{\B^{5/2}}\right)^2 = 
\tilde{A} \delta^{3/2} = -\delta^3 h_{xx}(1),  
\end{equation}
and thus we have determined the last constant,
\begin{equation}
\ti{A} = -\delta^{3/2} h_{xx}(1).
\end{equation}
Substitution of this value of $\tilde{A}$ into \eqref{H1aa} gives the \emph{effective} curvature to be applied as a boundary condition on the outer problem: $H_{xx}(1) = h_{xx}(1)$. 

In summary, in order to determine the leading-order outer solution in the $\B \to 0$ limit, we solve the outer boundary value problem
\begin{subequations} \label{bvp3all}
\begin{gather}
-\delta^3 H_{xxx} = \R(H; h_\infty), \label{bvp3} \\ 
H(0) = 1, \quad H_{xx}(0) = 0, \label{bvp3a} \\
H(1) = h(1), \quad H_x(1) = h_x(1), \quad H_{xx}(1) = h_{xx}(1), \label{bvp3b}
\end{gather}  
\end{subequations}
in the following way: first, a value of $\{h_\infty, H(1)\}$ is used as a starting guess. The Landau-Levich equation \eqref{LandauLevich} is solved from a large number, $x = x_\text{max}$, with $h$ near its far field height, and the integrator is stopped once $h = H(1)$. At this point, we solve the third-order boundary value problem \eqref{bvp3} subject to three (of the five) conditions of \eqref{bvp3a} and \eqref{bvp3b}. The shooting algorithm is then repeated to converge to correct values of $\{h_\infty, H(1)\}$ in order to satisfy the remaining two conditions. 

It is a curious fact that the effective boundary conditions on the outer problem, \eqref{bvp3all}, impose the constraint that the elastic plate, $H(x)$, and the free surface, $h(x)$, are continuous in their profiles, derivatives, and curvatures, and so together, behave very similarly to a continuous free-surface fluid. However, we should recall that the equation for $H(x)$ imposes a zero-slip condition on the free surface, whereas the velocity profiles for $h(x)$ will slip at the free boundary. Additionally, the interpretation of the elastic plate as continuously attaching to the fluid is only true of the \emph{outer} profile, as is clearly shown in Figure~\ref{fig:Hxxtau0}. In particular, there will indeed be a rapid variation in the curvature of the elastic plate from its non-zero effective value (proportional to the curvature $h''(x)$) and the zero value required for the moment-free constraint.

\subsubsection{Modification for $p_0 \neq 0$} \label{sec:p0notzero}

Now that we have understood the boundary layer procedure, we see that the introduction of Regions I and III is not strictly necessary in order to derive the effective boundary conditions on the outer problem. This is because the solution in Region II contains the information for all three boundary layer regions, and in fact, the boundary conditions exactly at $x = 1$ can be applied directly to II. The crucial quality that allows for this is that the terms that change balance between I, II, and III are exponential in nature.

With the introduction of $p_0 \neq 0$, another boundary layer (or rather, three new layers) must be introduced at $x = 0$. Using the same naming convention for the three regions, the solution in the intermediate region, which consists of the dominant balance $\B^5 H^{(5)} \sim \delta^3 H^{(3)}$, possesses the curvature 
\begin{equation}
H_{xx}^\text{II, left} \sim \frac{a_1}{\delta^{3/2}} \exp\left[-\frac{\delta^{3/2}}{\B^{5/2}} x\right] + a_2,
\end{equation}
where, like in Sec.~\ref{sec:region2}, we have scaled the constants $a_1$ and $a_2$ so that the curvature remains non-zero as $\B \to 0$. Next, matching with the innermost solution, II $\to$ III requires that $H_{xx}(0) = 0$ for the hinged plate. Thus $a_2 = -a_1/\delta^{3/2}$. The last step is to apply the pressure condition \eqref{pressurebc1}, requiring $\B^5 H^{(4)}(0) = p_0$; this gives $a_1 = p_0/\delta^{3/2}$, so the final intermediate solution is 
\begin{equation}
H_{xx}^\text{II, left} \sim \frac{p_0}{\delta^3} \left(\exp\left[-\frac{\delta^{3/2}}{\B^{5/2}} x\right] - 1\right),
\end{equation}
and we observe that the effective curvature to apply to the outer solution is $-p_0/\delta^3$. 

\begin{figure}\centering
\includegraphics{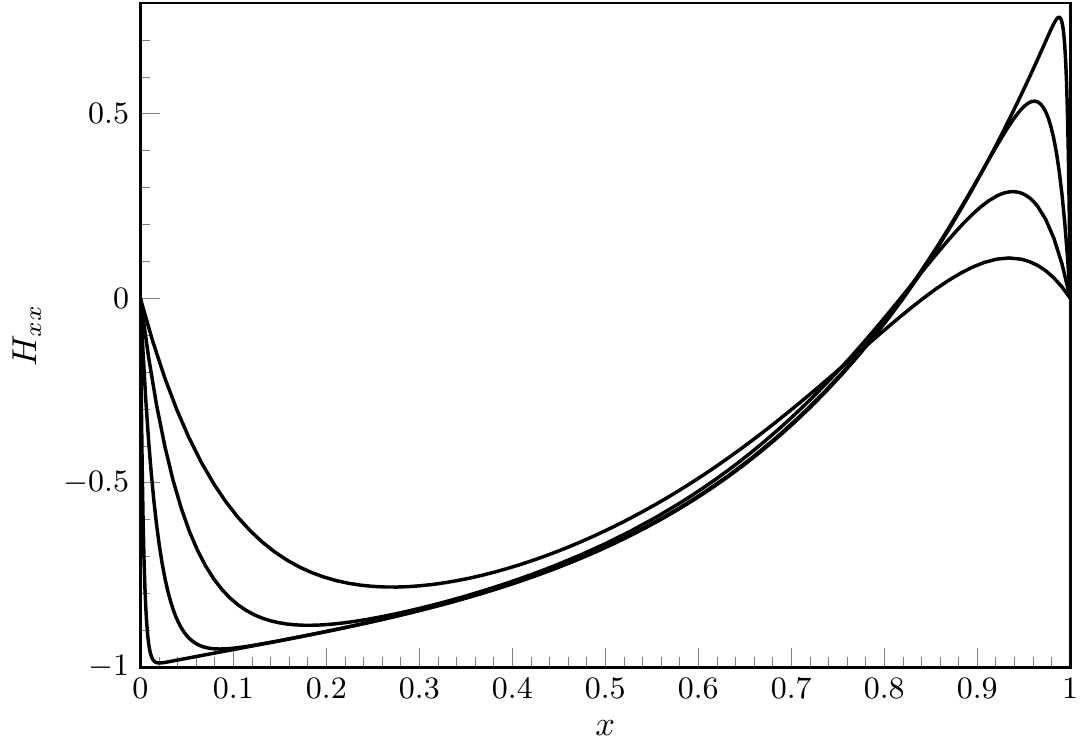}
\caption{ $H_{xx}$ for various values of $\tau$. For these solutions, $\delta = 1$ and $p_0 = 1$. From top to bottom measured near $x = 1$, the curves correspond to $\tau = 0.1, 0.2, 0.3, 0.4$. \label{fig:Hxxtau0_nonp0}}
\end{figure}

This completely determines the problem, and we are left with the effective outer problem:
\begin{subequations} \label{bvp3all_nonp0}
\begin{gather}
-\delta^3 H_{xxx} = \R(H; h_\infty), \label{pneq0a} \\ 
H(0) = 1, \quad H_{xx}(0) = -\frac{p_0}{\delta^3},  \\
H(1) = h(1), \quad H_x(1) = h_x(1), \quad H_{xx}(1) = h_{xx}(1), \label{pneq0caa}
\end{gather}  
\end{subequations}
which are solved similarly to the set \eqref{bvp3all}. Numerical solutions are shown in Figure~\ref{fig:Hxxtau0_nonp0}, and indeed we observe that the outer curvature $H_{xx}$ tends to non-zero constants at either end of the interval. 

\section{Asymptotic analysis of slow and fast plate speeds} \label{sec:deltlimit}

In this section, we examine the limits whereby the inverse capillary number, $\delta^3 = \gamma \epsilon^3/\mu U = 1/\Ca$ tends to zero or infinity. Based on the analysis for the rigid plate in \rigidpaper{}, we understand that in the limit $\delta \to \infty$, this corresponds to very slow motion of the bottom substrate where both the plate and the fluid uniformly tends towards the configuration with uniform height. In contrast, the limit of $\delta \to 0$ encapsulates the situation of a very fast motion of the bottom substrate, which causes the fluid to be uniform nearly everywhere except within a boundary layer at $x = 0$. In this section, we find that this description of the equilibrium configuration still holds for flexible plates, $\B \neq 0$, but like in Sec.~\ref{sec:Blimit}, the elastic effects will introduce additional boundary layers near the plate edges, which are crucial for the matching process (and the eventual determination of the far field height, $h_\infty$). 

\subsection{The limit $\delta \to \infty$ (slow substrate motion and/or strong surface tension)} \label{sec:deltinf}

In the limit that $\delta \to \infty$, it can be verified that both the plate, $H$, and free-surface, $h$ uniformly tend to unity. We then expand
\begin{equation} \label{deltinf:series}
 H(x) = 1 + \sum_{n=1}^\infty \frac{H_n}{\delta^n},
  \end{equation}
%

Like in the analysis for variable $\B$, we are interested in developing a uniformly valid solution for the plate curvature, $H_{xx}$, which is expected to be composed of two inner regions near the boundaries, $H_{xx}^\lefty$ and $H_{xx}^\righty$, and an outer region, $H_{xx}^\outie$. Also like the previous analysis, if the reservoir pressure is zero, the boundary layer on the left may not be necessary.  Let us consider the case of $p_0 \neq 0$. 

We first consider the outer region. Using $H$ and $h_\infty \sim 1$ and assuming a dominant balance between the surface tension \mycirc{2} and pressure term \mycirc{3}, we find that $\delta^3 H^{(3)} \sim 6$ or 
\begin{equation} \label{deltinf:Hxx}
H^{\outie}_{xxx} \sim \frac{1}{\delta^3} (6x - d),
\end{equation}
for constant $d$, and where the outer region is restricted to fixed values of $x \in (0, 1)$. Notice further that we no longer assume that the curvature is $\Oh(1)$ in the asymptotic limit, as we had done in \eqref{keyassume}. 

In the left boundary layer, we seek a balance between the elasticity \mycirc{1} and surface tension \mycirc{2} terms of \eqref{elasticbalance}, and thus require a boundary layer of size $x = \Oh(\delta^{-3/2})$. Solving $\B^5 H^{(5)} \sim \delta^3 H^{(3)}$, and using the moment free condition, $H_{xx}(0) = 0$, we obtain 
\begin{equation}
H_{xx}^\lefty \sim -\frac{D}{\delta^3} \left[ 1 - \exp\left( -\frac{\delta^{3/2} x}{\B^{5/2}} \right) \right],
\end{equation}
for a constant, $D$. In order for this curvature to match the outer curvature \eqref{deltinf:Hxx}, we take $x\delta^{3/2} \to \infty$ and thus require $D = d$. Moreover, imposing the pressure condition \eqref{pressurebc1} yields $D = p_0$. 

It remains to determine the curvature, $H_{xx}^\righty$, near $x = 1$. We introduce a boundary layer of size $x = 1 - \Oh(1/\delta^{3/2})$ and balance the elasticity \mycirc{1} and surface tension \mycirc{2} terms of \eqref{elasticbalance}. Imposing the condition that $H_{xx}(1) = 0$ yields
\begin{equation}
H_{xx}^\righty  \sim \frac{G}{\delta^3}\left[ 1 - \exp\left(-\frac{\delta^{3/2}(1-x)}{\B^{5/2}}\right)\right],
\end{equation}
where the constant $G = (6 - d) = (6 - p_0)$ once $H_{xx}^\righty$ is matched to $H_{xx}^\outie$. Thus, we conclude altogether that 
\begin{equation} \label{deltinf:solH}
H_{xx} \sim \begin{cases}
-\dfrac{p_0}{\delta^3} \left[ 1 - \exp \left(-\dfrac{x \delta^{3/2}}{\B^{5/2}}\right)\right], & \text{for $x = \Oh(\delta^{3/2})$}, \\
\dfrac{1}{\delta^3} (6x - p_0), & \text{for $x \in (0, 1)$}, \\
\dfrac{6-p_0}{\delta^3} \left\{ 1- \exp \left(-\dfrac{(1-x)\delta^{3/2}}{\B^{5/2}}\right)\right\}, & \text{for $x = 1 - \Oh(\delta^{-3/2})$}.
\end{cases}
\end{equation}

\begin{figure}\centering
\includegraphics[width=1.0\textwidth]{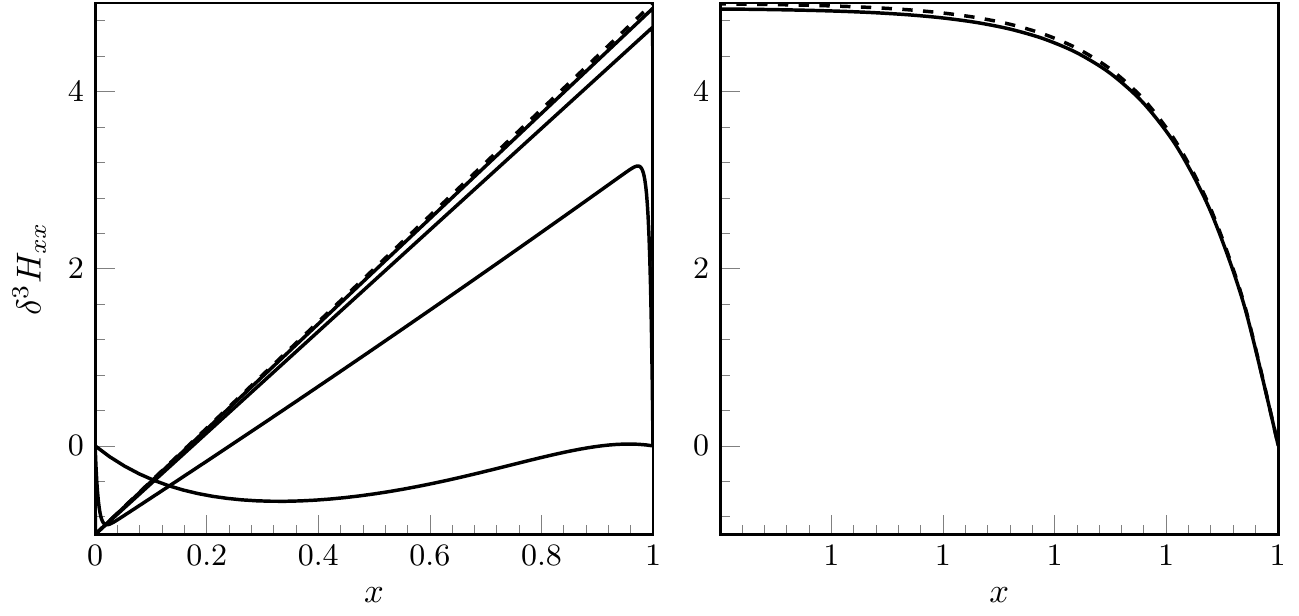}
\caption{$\delta^3 H_{xx}$ for $\tau = 0.5$, $p_0 = 1$. The curves correspond to (from top to bottom at the right) $\delta = 500, 10, 10, 1$. On the right is an enlargement of the region near $x = 1$ for the curve $\delta = 500$. The asymptotic prediction is shown dashed.\label{Hxxdeltinf}}
\end{figure}

The asymptotic solutions are shown against the numerical solutions in Figure \ref{Hxxdeltinf} for values of $\delta = 1, 10, 100, 800$. We have plotted $\delta^3 H_{xx}$ in order to remove the algebraic dependence on $\delta$. The dashed profiles are formed by combining the asymptotic predictions in the three regions.

With the curvature of the plate entirely determined, we are free to proceed similarly to \rigidpaper{} and study the Landau-Levich equation \eqref{LandauLevich} in the limit of $\delta \to \infty$. In this limit, the deflection from $h \sim 1$ for finite $x$ is exponentially small in $\delta$, so it is better to seek a re-scaling near the plate edge, $x = 1$. Setting $x = 1 + \delta X$, we substitute
\begin{equation}
  h(x) = 1 + \sum_{n=1}^\infty \frac{h_n(X)}{\delta^n} \quad \text{and} \quad
 h_\infty = 1 + \sum_{n=1}^\infty \frac{h_{\infty n}}{\delta^n}
\end{equation}
into the re-scaled Landau-Levich equation
\begin{equation}
  h_{XXX} = \frac{3(h_\infty - h)}{h^3}.
\end{equation}

Solving the equation at $\Oh(1/\delta)$, we set two integration constants to zero to remove the exponentially growing modes (see [8,9] in Table~\ref{tab:pinned}), and match the free surface to the plate $h_1(X = 0) = H_1(x = 1)$ where $H_1$ is from the series expansion \eqref{deltinf:series}. Written in terms of $x$, this yields
\begin{equation} \label{deltinf:h1}
h_1 = h_{\infty 1} + \Bigl[ H_1(1) - h_{\infty 1}\Bigr] \exp\left[-\frac{3^{1/3} (x-1)}{\delta}\right].
\end{equation}

\begin{figure}\centering
\includegraphics{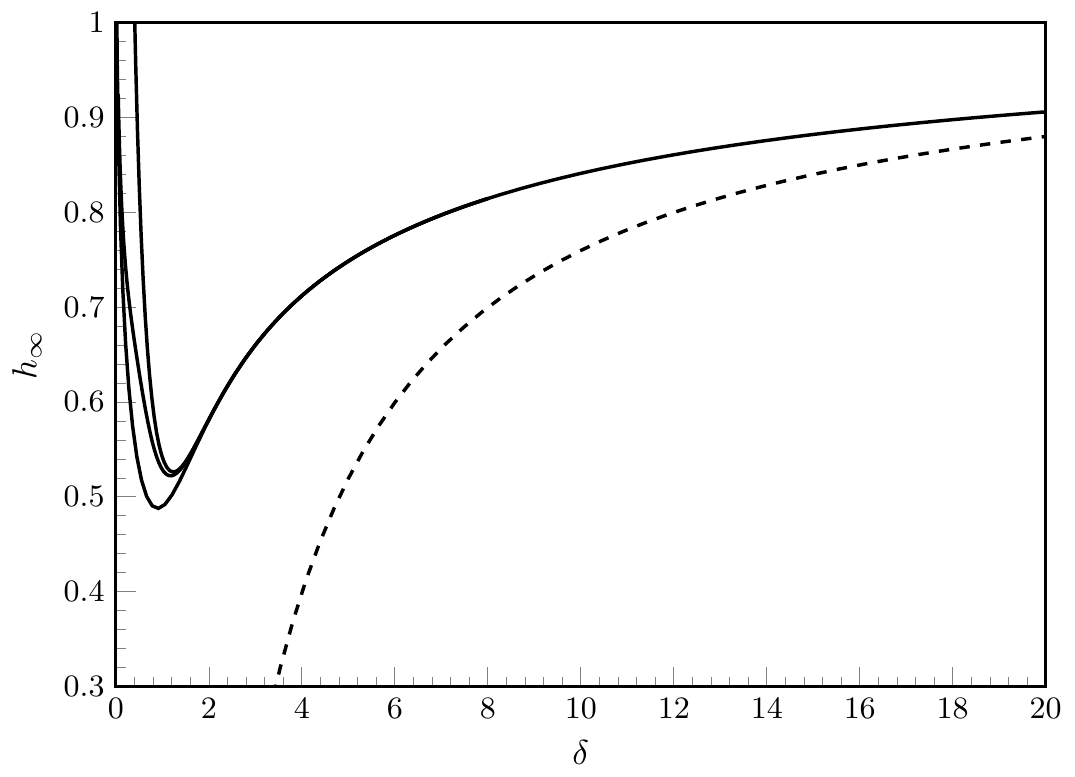}
\caption{Numerical values of the far field height, $h_\infty$, plotted as a function of the elasticity parameter, $\B$ for $p_0 = 1$. The solid curves are, from top to bottom, $\B= 0.2, 0.5, 1$. The two-term asymptotic approximation for $h_\infty$ in the limit $\delta \to \infty$ from \eqref{deltinf:hinf} is shown dashed. \label{hinf_deltinfty}}
\end{figure}

However, note that comparing the expanded form \eqref{deltinf:series} to the solution \eqref{deltinf:solH}, we observe that the $\Oh(1/\delta)$ contribution to the plate is identically zero $H_1 \equiv 0$, and thus the free-surface correction in \eqref{deltinf:h1} only depends on $h_{\infty 1}$. Consequently, \eqref{deltinf:h1} yields an expression for the leading-order free-surface curvature near the plate edge, given by $h_{xx}(1) \sim - 3^{2/3} h_{\infty 1}/\delta^3$. The normal load at the edge of the elastic plate is $\B^5 H^{(5)}(1) \sim (p_0 - 6)/\delta^3$, which follows from \eqref{deltinf:solH}. Finally, the fluid pressure and plate load are related through \eqref{pressurebc2}. Solving for the far field height gives
\begin{equation} \label{deltinf:hinf}
h_\infty = 1 + \left[\frac{p_0 - 6}{3^{2/3}}\right] \frac{1}{\delta},
\end{equation}
which is verified in Figure \ref{hinf_deltinfty}.

\subsection{The limit $\delta \to 0$ (fast substrate motion and/or weak surface tension)} \label{sec:delt0}

In the limit $\delta \to 0$, the plate uniformly tends to unit height, $H \sim 1$, while the downstream fluid tends to $h \sim h_{\infty} \sim 1/2$, except near the plate edge where the free surface rises rapidly to match the plate. In terms of the analysis for the elastic plate, the principal difference between the $\delta \to 0$ limit here, and $\delta \to \infty$ of Sec.~\ref{sec:deltinf} and $\B \to 0$ limit of Sec.~\ref{sec:B0} is that for the case here, the $\delta \to 0$ limit is regular, and there are no boundary layers in the plate. 

We can then expand
\begin{equation}
  H(x) = 1 + \sum_{n=1}^\infty \delta^n H_n \quad \text{and} \quad
   h_\infty = \frac{1}{2} + \sum_{n=1}^\infty \delta^n h_{\infty n}.
\end{equation}
At $\Oh(\delta)$ in \eqref{elasticeq}, we find
\begin{equation}
\B^5 H_1^{(5)}  = 6(H_1 - 2h_{\infty 1}),
\end{equation}
which possesses the general solution 
\begin{subequations} \label{eq:delt0_gen}
\begin{equation}
H_1(x; \, h_{\infty 1}) = 2h_{\infty 1} + \sum_{j=0}^4 C_j \exp \left[ \frac{e^{2\pi \im k/5} 6^{1/5} x}{\B}\right],
\end{equation}
for constants $C_j$. The six values $\{C_0, \ldots, C_4, h_{\infty 1}\}$ can be solved by imposing the six limiting boundary conditions 
\begin{alignat}{3}
H_1(0) &= 0, \quad H_1''(0) &&= 0, \quad H_1'''(0) &&= \frac{p_0}{\B^5}, \\ 
H_1''(1) &= 0, \quad H_1'''(1) &&= 0, \quad H_1''''(1) &&= -\frac{h_{0xx}(1)}{\B^5}.
\end{alignat}
\end{subequations}

The only unknown in the above set of boundary conditions is the value of $h_{0xx}(1)$. This value can be found by numerically solving the Landau-Levich equation \eqref{LandauLevich} beginning with initial conditions at large values of $x$, and stopping once $h(1) \sim h_0(1) = 1$ is reached. Based on the results in \rigidpaper, the value is approximately
\begin{equation}
h_{0xx}(1) \sim \frac{1.7639}{\delta^2}.
\end{equation}

\begin{figure}\centering
\includegraphics[width=1.0\textwidth]{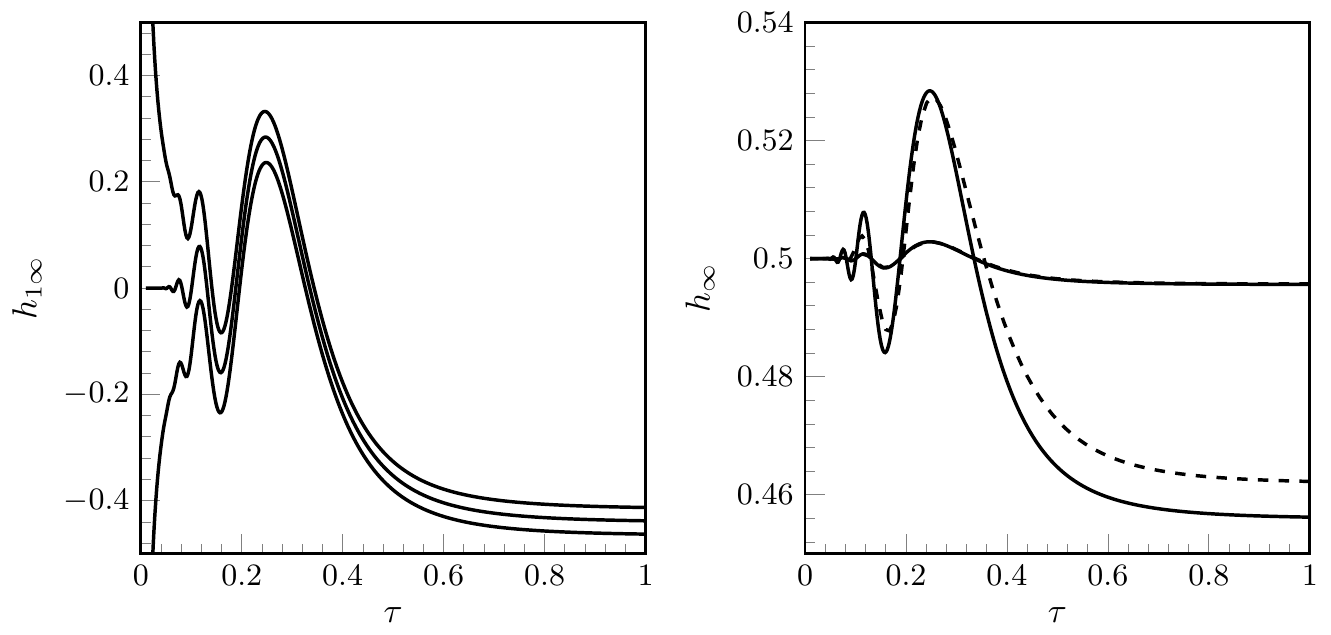}
\caption{(Left) Plot of $h_{\infty 1}$ versus $\B$ at $p_0 = \{0.1, 0, -0.1\}$ from top to bottom; (right) plot of $h_\infty$ versus $\B$ at $p_0 = 0$ for asymptotic solutions (solid) and full numerical solutions (dashed) at $\delta = 0.1$ (large curves) and $\delta = 0.01$ (flat curves). The two two curves for $\delta = 0.01$ are nearly indistinguishable. \label{fig:hinf_delt0}}
\end{figure}

In the left frame of Figure \ref{fig:hinf_delt0}, we display the asymptotic values of $h_{1\infty}$ for $p_0 = \{0, 0.1, -0.1\}$, calculated using the above approach. Note that the solution that corresponds to $p_0 = 0$ looks to be regular as $\B \to 0$, but solutions for non-zero pressure, $p_0$ diverge in the same limit, and this was similarly observed in the previous work \rigidpaper{}. In the right frame of Figure \ref{fig:hinf_delt0}, we compare the asymptotic $\delta \to 0$ two-term prediction using $h_\infty \sim 1/2 + \delta h_{\infty 1}$ with the full numerical solution at $\delta = 0.1$ and $\delta = 0.01$ and for pressure $p_0 = 0$. It is confirmed that the fit is very good, and by $\delta = 0.01$, the curves are nearly visual indistinguishable on the graph. 

The one regime we were not able to resolve concerns the increasingly rapid oscillations that are observed in Figure~\ref{fig:hinf_delt0} when both $\delta \to 0$ and $\B \to 0$. In the analysis leading to asymptotic prediction \eqref{eq:delt0_gen}, $\B$ was assumed to be fixed while $\delta \to 0$. However,  we note that the solutions that correspond to (afterwards) setting $\B \to 0$ have $H_1(1) \to -\infty$, and are thus inadmissible as soon as $H_1(1)$ exceeds $\Oh(1/\delta)$ in magnitude. This suggests there exists a distinguished limit with $\delta = \B \to 0$. 

\begin{figure}\centering
\includegraphics{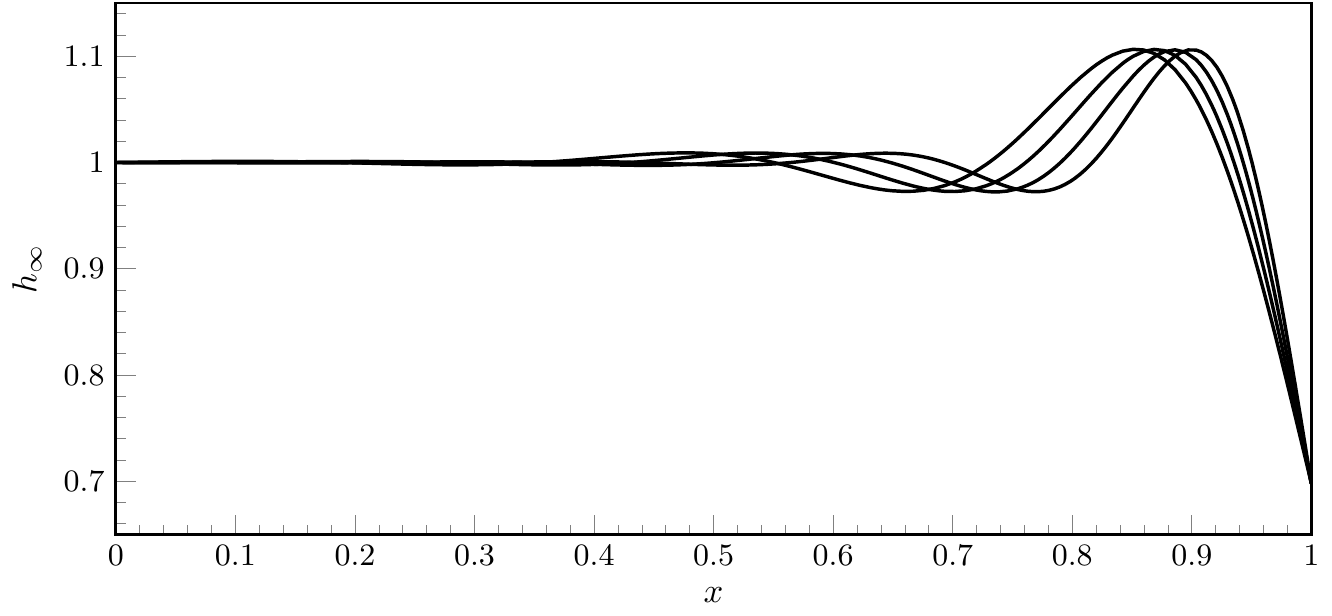}
\caption{Plate profiles, $H(x)$ for $p_0 = 0$, $\delta = \B$, and $\B = 0.0500, 0.0573, 0.0652, 0.0735$ from right to left. There is a distinguished limit as $\delta, \B \to 0$. \label{fig:smalldelttau}}
\end{figure}

Example of solutions in this limit are given in Figure~\ref{fig:smalldelttau} for the case of $p_0 = 0$, and give values of $\B$. We observe is that as $\delta, \B \to 0$, $H(x)$ exhibits a series of maxima and minima, for which the main maximum moves closer towards the boundary at $x = 1$ in the singular limit. The behaviour of these ripples seems to parallel those observed in the analysis of \cite{wilson_1983}, \cite{snoeijer_2008} and \cite{benilov_2010}, but we shall leave this particular special limit as an open problem.

\section{Discussion}

We have presented an asymptotic and numerical analysis of a mathematical model for a pinned elastic plate interacting with a thin viscous fluid film, as the system is forced by a moving bottom substrate. Four asymptotic limits were studied: the limits of large and small plate elasticity, and the limits of small and large Capillary numbers. With the exception of small elasticity, $\B \to \infty$, where the plate tends to a rigid solid, the remaining three limits require matched asymptotic expansions order to perform incorporate the singular effects at the edges of the plate. The emergence of such boundary layers are often the culprit for non-convergence of numerical solutions, and we expect that this behaviour is generic in studies involving highly flexible plates interacting with thin films. 

We note that there are numerous connections between the third-order Landau-Levich equation \eqref{LandauLevich}, which is commonly encountered in the modeling of coating flows, and what we introduced as a fifth-order Landau-Levich-\emph{like} equation (effectively a beam equation but with Reynolds equation embedded within to provide the fluid flux and pressure). Thus, many of the asymptotic and numerical properties that are used to study the Landau-Levich equation share a correspondence with the properties of the elastic equation. A clear connection, for instance, exists between the development of an apparent curvature in the $\B \to 0$ and $\delta \to \infty$ asymptotics for the outer solution (see Figures~\ref{fig:Hxxtau0}--\ref{Hxxdeltinf}) and the analysis of moving contact lines, where the goal is to develop an apparent or macroscopic contact angle \citep{dussan_1974, hocking_1981, ren_2014}, related to an inner region near the singularity.

In regards to further research on the particular pinned plate problem presented in this paper, we note that it remains an open problem to characterize the global bifurcation structure of the solutions. As noted in the introduction, the nonlinearity of the associated differential equations is often accompanied by non-uniqueness of the solutions, and the existence or non-existence solutions was discussed in \cite{moriarty_1996}, and also the rigid analysis of our previous work \rigidpaper{}. Rigorous results can be established for the case of a rigid plate (as was done in \citealt{mcleod_1996}), but for the convoluted boundary-value system we have presented here, the analogous theory remains unresolved. The non-uniqueness of the fifth-order Landau-Levich-\emph{like} beam equation (though uncoupled from another thin-film equation) was also noted in the work of \cite{dixit_2013}.

\begin{figure}\centering
\vspace*{0.1cm}
\includegraphics{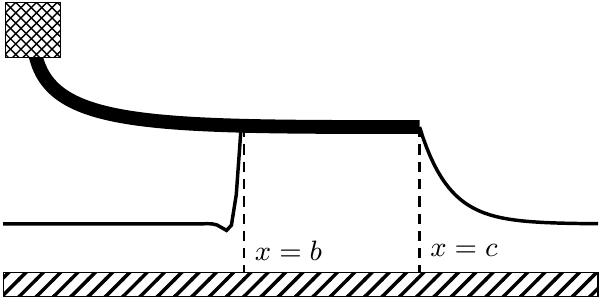} \hfill
\includegraphics{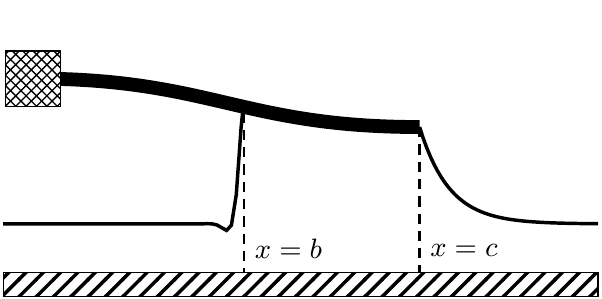} 
\caption{Two example flows with free contact points, $x = a$ and $x = b$, which must be determined as part of the solution. Vertically clamped plate (left) and horizontally clamped plate (right), held stationary while the bottom substrate is moved at a constant speed. \label{fig:sketch}}
\end{figure}

There are two interesting directions we highlight for more general future work. The first variation of our problem is inspired by the work of \cite{seiwert_2013}, who studied the problem of a vertically clamped plate, used within the blade-coating system illustrated in Figure \ref{fig:sketch}a. For such geometries, it may not be possible to assume a weakly deflected plate from the horizontal ($x$-axis), but linear plate theory can still used as a function of the plate's arclength. Alternatively, one can consider (Figure~\ref{fig:sketch}b) a configuration of a horizontally clamped plate, which will be more similar to the theory we have presented in this paper. Note that in order to preserve the applicability of lubrication theory, it must be assumed that the plate is sufficiently flat during contact. Moreover, a particular challenge in studying such configurations is due to the upstream connection point being \emph{a priori} unknown and having to be solved as part of the solution. 

The second direction we highlight for future work concerns an extension of the pinned-plate problem to the study of plate-fluid interactions where the underlying substrate is curved. This direction has a great deal in common with the contact lens models that had motivated the freely-floating blade problem in our earlier work \rigidpaper{}. The theory of thin film flows on curved manifolds is not as well developed as for flat surfaces [though \emph{c.f.} \cite{myers_2002, howell_2003, jensen_2004, trinh_2014b}], but we have been intent on studying the effects of variable substrate geometry on the type of elastocapillary flows studied in this paper. Work on this topic is ongoing.

\bigskip\noindent\emph{Acknowledgements:} We would like thank Drs. Peter Howell (Oxford) and Dominic Vella (Oxford) for valuable discussions during the course of this work. PHT thanks Lincoln College (Oxford) for financial support. SKW is presently a Leverhulme Trust Research Fellow (2013--2015) supported by award RF-2013-355.

\providecommand{\noopsort}[1]{}

\end{document}